\documentclass{article}
\usepackage[utf8]{inputenc}
\usepackage{amsmath}
\usepackage{amssymb}
\usepackage{graphicx}
\usepackage{hyperref}
\usepackage{booktabs}
\usepackage{siunitx}
\usepackage{tikz}
\usetikzlibrary{shapes.geometric}

\newcommand{\bx}{\mathbf{x}}
\newcommand{\bz}{\mathbf{z}}
\newcommand{\bw}{\mathbf{w}}
\newcommand{\bb}{\mathbf{b}}
\newcommand{\N}{\mathcal{N}}

\newcommand{\Embu}{P}
\newcommand{\Embi}{Q}
\newcommand{\embu}{\mathbf{p}}
\newcommand{\embi}{\mathbf{q}}
\newcommand{\sembu}{p}
\newcommand{\sembi}{q}
\renewcommand{\O}{\mathcal{O}}

\author{
\begin{tabular}{ccc}
  Steffen Rendle\thanks{Google Research, Mountain View, USA. \{srendle,walidk,liqzhang,janders\}@google.com}
  &&
  Walid Krichene\footnotemark[1]   \vspace*{0.2cm}
\\
  Li Zhang\footnotemark[1] &&
  John Anderson\footnotemark[1]
\end{tabular}
}

\title{Neural Collaborative Filtering vs. \\Matrix Factorization Revisited}

\date{}

\begin{document}

\maketitle

\begin{abstract}
Embedding based models have been the state of the art in collaborative filtering for over a decade.
Traditionally, the dot product or higher order equivalents have been used to combine two or more embeddings, e.g., most notably in matrix factorization.
In recent years, it was suggested to replace the dot product with a learned similarity e.g. using a multilayer perceptron (MLP).
This approach is often referred to as \emph{neural collaborative filtering} (NCF).
In this work, we revisit the experiments of the NCF paper that popularized learned similarities using MLPs.
First, we show that with a proper hyperparameter selection, a simple dot product substantially outperforms the proposed learned similarities.
Second, while a MLP can in theory approximate any function, we show that it is non-trivial to learn a dot product with an MLP.
Finally, we discuss practical issues that arise when applying MLP based similarities and show that MLPs are too costly to use for item recommendation in production environments while dot products allow to apply very efficient retrieval algorithms.
We conclude that MLPs should be used with care as embedding combiner and that dot products might be a better default choice.
\end{abstract}

\section{Introduction}

Embedding based models have been the state of the art in collaborative filtering for over a decade.
A core operation of most of these embedding based models is to combine two or more embeddings.
For example, combining a user embedding with an item embedding to obtain a single score that indicates the preference of the user for the item. This can be viewed as a \emph{similarity function} in the embedding space.
Traditionally, a dot product or higher order products have been used for the similarity.
Recently, it has become popular to learn the similarity function with a neural network.
Most commonly, a multilayer perceptron (MLP) is used for the network architecture~(e.g. \cite{hu:kdd18, zamani:wsdm20, zhao:wsdm20, jawarneh:20, jiarui:20, mattson2019:mlperf}).
This approach is often referred to as \emph{neural collaborative filtering} (NCF)~\cite{he:www17}.
The rationale is that MLPs are general function approximators so that they should be strictly better than a fixed similarity function such as the dot product.
This has made NCF the model of choice for comparison in many recommender studies~(e.g. \cite{hu:kdd18, zamani:wsdm20, wei:wsdm18, zhao:wsdm20, jawarneh:20, mattson2019:mlperf}).

In this work, we study MLP versus dot product similarities in more detail.
We start with revisiting the experiments of the NCF paper~\cite{he:www17} that popularized the use of MLPs in recommender systems.
We show that a carefully configured dot product baseline largely outperforms the MLP.
At first glance, it looks surprising that the MLP, which is a universal function approximator, does not perform at least as well as the dot product.
We investigate this issue in a second experiment and show empirically that learning a dot product with high accuracy for a decently large embedding dimension requires a large model capacity as well as many training data.
Besides prediction quality, we also discuss the inference cost of dot product versus MLPs, where dot products have a large advantage due to the existence of efficient maximum inner product search algorithms.
Finally, we discuss that dot product vs MLP is not a question of whether a deep neural network (DNN) is useful.
In fact, many of the most competitive DNN models, such as transformers in natural language processing~\cite{devlin:arxiv18} or resnets for image classification~\cite{he:cvpr16}, use a dot product similarity in their output layer.

To summarize, this paper argues that MLP-based similarities for combining embeddings should be used with care.
While MLPs can approximate any continuous function, their inductive bias might not be well suited for a similarity measure.
Unless the dataset is large or the embedding dimension is very small, a dot product is likely a better choice.

\section{Definitions}
\label{sec:def}

In this section, we formalize the problem and review dot product (esp., matrix factorization) and learned similarity functions (esp., MLP and NeuMF).
We denote matrices by upper case letters $X$, vectors by lowercase bold letters $\bx$, scalars by lowercase letters $x$.
A concatenation of two vectors $\bx, \bz$ is denoted by $[\bx,\bz]$.

Our paper studies functions $\phi : \mathbb{R}^d \times \mathbb{R}^d \to \mathbb{R}$ that combine two $d$-dimensional embedding vectors $\embu \in \mathbb{R}^d$ and $\embi \in \mathbb{R}^d$ into a single score.
For example $\embu$ could be the embedding of a user, $\embi$ the embedding of an item, and $\phi(\embu, \embi)$ is the affinity of this user to the item.

The embeddings $\embu$ and $\embi$ can be model parameters such as in matrix factorization, but they can also be functions of other features, for example the user embedding $\embu$ could be the output of a deep neural network taking user features as input. From here on, we focus mainly on the similarity function $\phi$ but in Section~\ref{sec:dnn} we will discuss the embeddings in more detail.

\begin{figure}
    \centering

\newcommand\tower[3]{
\draw[rounded corners=5pt] (#1+1,4) rectangle ++(#2-2,0.6) node[pos=.5] {#3};
\node [trapezium,trapezium angle=80,inner xsep=0pt,minimum height=1.5cm,text width=.3cm,draw] at (#1+#2/2, 3) {};
}

\resizebox{\textwidth}{!}{%
\begin{tikzpicture}[]
\large
\tower{1}{4}{$\embu \in \mathbb R^d$}
\tower{4}{4}{$\embi \in \mathbb R^d$}
\node[minimum size=15pt,inner sep=0pt] (dot) at (4.5, 6) {$\phi^{\text{dot}}(\embu, \embi) = \langle \embu, \embi \rangle$};
\node[circle, draw, minimum size=15pt,inner sep=0pt] (dot) at (4.5, 5.2) {$\cdot$};
\draw (3, 4.6) -- (dot);
\draw (6, 4.6) -- (dot);
\end{tikzpicture}
\hspace{1cm}
\begin{tikzpicture}[]
\tower{1}{4}{$\embu \in \mathbb R^d$}
\tower{4}{4}{$\embi \in \mathbb R^d$}
\node[minimum size=15pt,inner sep=0pt] (dot) at (4.5, 6) {$\phi^{\text{MLP}}(\embu, \embi) = \textbf{f}_{W_l,\bb_l}(\ldots \textbf{f}_{W_1,\bb_1}([\embu, \embi])\ldots)$};
\node [trapezium,trapezium angle=22,inner xsep=0pt,minimum height=.8cm,text width=.6cm,draw] at (4.5, 5.2) {MLP};
\end{tikzpicture}
}
\caption{A model with dot product similarity (left) and MLP-based learned similarity (right).}
\label{fig:similarity_fn}
\end{figure}
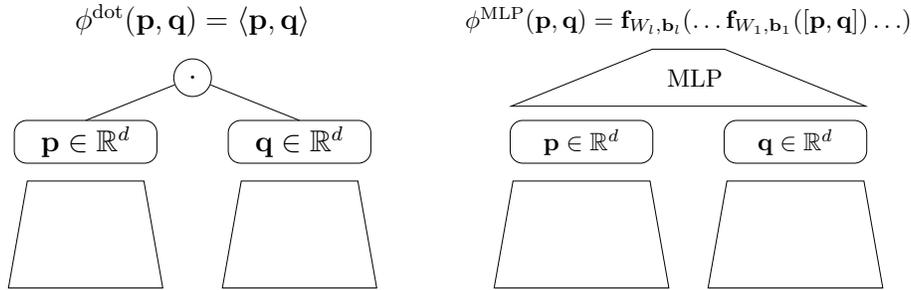

\paragraph{Dot Product}

The most common combination of two embeddings is the dot product.
\begin{align}
    \phi^{\text{dot}}(\embu, \embi) := \langle \embu, \embi \rangle = \embu^T \embi = \sum_{f=1}^d \sembu_f \sembi_f .
\end{align}
If $\embu$ and $\embi$ are free model parameters, then this is equivalent to matrix factorization. A common trick is to add explicit biases:
\begin{align}
    \phi^{\text{dot}}(\embu, \embi) := b + \sembu_{1} + \sembi_{1} + \langle \embu_{[2,\ldots,d]}, \embi_{[2,\ldots,d]} \rangle. \label{eq:mfbias}
\end{align}
This modification does not add expressiveness but has been found to be useful in many studies, likely because its inductive bias is better suited to the problem~\cite{paterek2007improving,koren:rshb11}.

\paragraph{Learned Similarity}

Multi layer perceptrons (MLPs) are known to be universal approximators that can approximate any continuous function on a compact set as long as the MLP has enough hidden states~\cite{cybenko1989approximation}.
One layer of a multi layer perceptron can be defined as a function $\textbf{f} : \mathbb{R}^{d_{\text{in}}} \rightarrow \mathbb{R}^{d_{\text{out}}}$:
\begin{align}
    \textbf{f}_{W,\bb}(\bx) = \boldsymbol{\sigma}(W \, \bx + \bb), \quad \boldsymbol{\sigma}(\bz) = [\sigma(z_1), \ldots, \sigma(z_\text{out})],
\end{align}
which is parameterized by $W \in \mathbb{R}^{\text{in} \times \text{out}}$, $\bb \in \mathbb{R}^{\text{out}}$ and an activation function $\sigma : \mathbb{R} \rightarrow \mathbb{R}$.
In a multilayer perceptron (MLP), several layers of $\textbf{f}$ are stacked, e.g., for a three layer MLP, $\textbf{f}_{W_3,\bb_3}(\textbf{f}_{W_2,\bb_2}(\textbf{f}_{W_1,\bb_1}(\bx)))$.

He et al.~\cite{he:www17} propose to replace the dot product with learned similarity functions for collaborative filtering.
They suggest to concatenate the two embeddings, $\embu$ and $\embi$, and apply an MLP:
\begin{align}
    \phi^{\text{MLP}}(\embu, \embi) := \textbf{f}_{W_l,\bb_l}(\ldots \textbf{f}_{W_1,\bb_1}([\embu, \embi])\ldots).
\end{align}
They further suggest a variation that combines the MLP with a weighted dot product model and name it \emph{neural matrix factorization} (NeuMF):
\begin{align}
\label{eq:ncf}
    \phi^{\text{NeuMF}}(\embu, \embi) := \phi^{\text{MLP}}(\embu_{[1, \ldots j]}, \embi_{[1 \ldots j}]) + \phi^{\text{GMF}}(\embu_{[j+1 \ldots d]}, \embi_{[j+1 \ldots d]}),
\end{align}
where GMF is a `generalized' matrix factorization model:
\begin{align}
    \phi^{\text{GMF}}(\embu, \embi) := \sigma( \bw^T  (\embu \odot \embi)) = \sigma(\langle \bw \odot \embu, \embi \rangle) = \sigma\left(\sum_{f=1}^d w_f \sembu_f \sembi_f\right). \label{eq:gmf}
\end{align}
with learned weights $\bw \in \mathbb{R}^d$.
For NeuMF, they recommend to use one part of the embedding (here the first $j$ entries) in the MLP and the remaining $d-j$ entries with the GMF.

Fig.~\ref{fig:similarity_fn} illustrates two models with dot product and MLP-based similarity.

\section{Revisiting NCF Experiments}

In this section, we revisit the experiments of the NCF paper~\cite{he:www17} that popularized the use of MLPs as embedding combiners in recommender systems.
We show that a simple dot product yields better results.

\subsection{Experimental setup}

The NCF paper \cite{he:www17} evaluates on an item retrieval task on two datasets: a binarized version of Movielens 1M~\cite{harper:15} and a dataset from Pinterest~\cite{geng:iccv15}.
Both are implicit feedback datasets, i.e. they contain only binary positive tuples between a user and an item.
For each user, the last item is held out and used as the test set, the remaining items of the user are placed into the training set.
For evaluation, each recommender ranks, for each user, a set of 101 items consisting of the withheld test item together with 100 random items.
For each user, the position at which the withheld item is ranked by the recommender is recorded, then two metrics are measured: (1)~Hit Ratio (i.e. Recall) among the top 10 ranked items -- which in this case is 1 if the withheld item is in the top 10 or 0 otherwise. (2)~NDCG among the top 10 ranked items -- which in this case is $1/log(r+1)$ where $r$ is the rank of the withheld item.
The average metric over all users is reported.
The authors have published the dataset splits and the evaluation code.
This allows us to evaluate on exactly the same setting and to compare our results directly with the ones reported in \cite{he:www17}.

\subsection{Models, loss and training algorithm}
We compare three models: MLP-learned similarity models introduced in~\cite{he:www17}, which use $\phi^{\text{MLP}}$ and $\phi^{\text{NeuMF}}$ respectively, and a simple matrix factorization baseline which uses $\phi^{\text{dot}}$ from Eq.~\eqref{eq:mfbias}. The only difference between these models is the similarity function. In particular, the embeddings~$\embu, \embi$ are free parameters in all models.
We train the matrix factorization baseline by minimizing a logistic loss with L2 regularization, using stochastic gradient descent (with no batching, no momentum or other variations) with negative sampling, as in the original paper~\cite{he:www17}\footnote{It is possible that a different loss or a different sampling strategy could lead to an even better performance of our method. However, we wanted to use the same loss and sampling strategy for all competing methods to ensure that this is a meaningful comparison, which will allow us to attribute differences in quality to the choice of similarity functions.}.
More precisely, for each training example (consisting of a user and a positive item), we sample $m$ negative items, uniformly at random.
Finally, we vary the embedding dimension $d \in \{16,32,64,96, 128, 192\}$.
Additional details about the setup can be found in Appendix~\ref{sec:ncf_exp_details}.

\subsection{Results}
\label{sec:ncf_exp}

\begin{figure}
    \centering
    \includegraphics[width=0.49\textwidth]{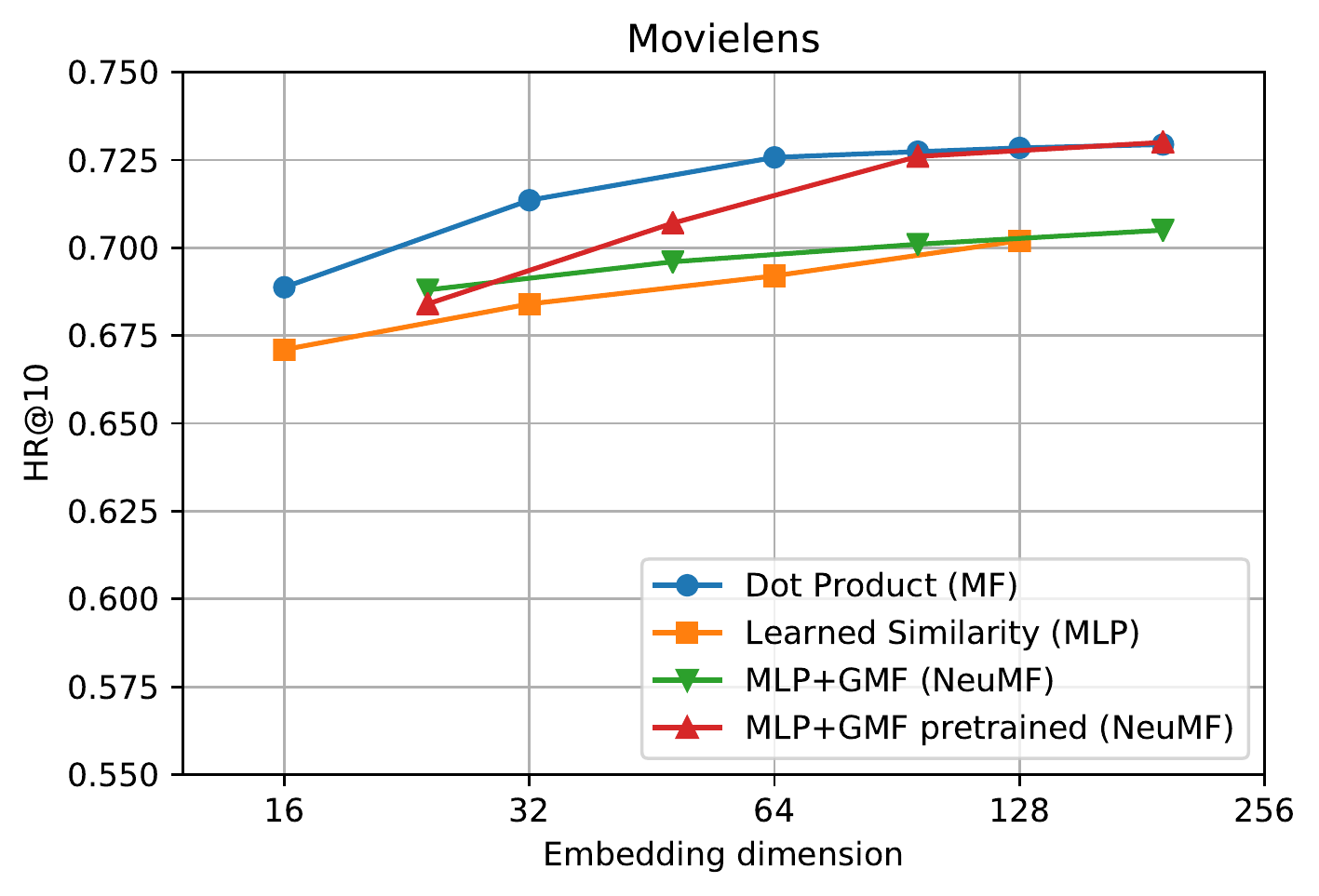}%
    \includegraphics[width=0.49\textwidth]{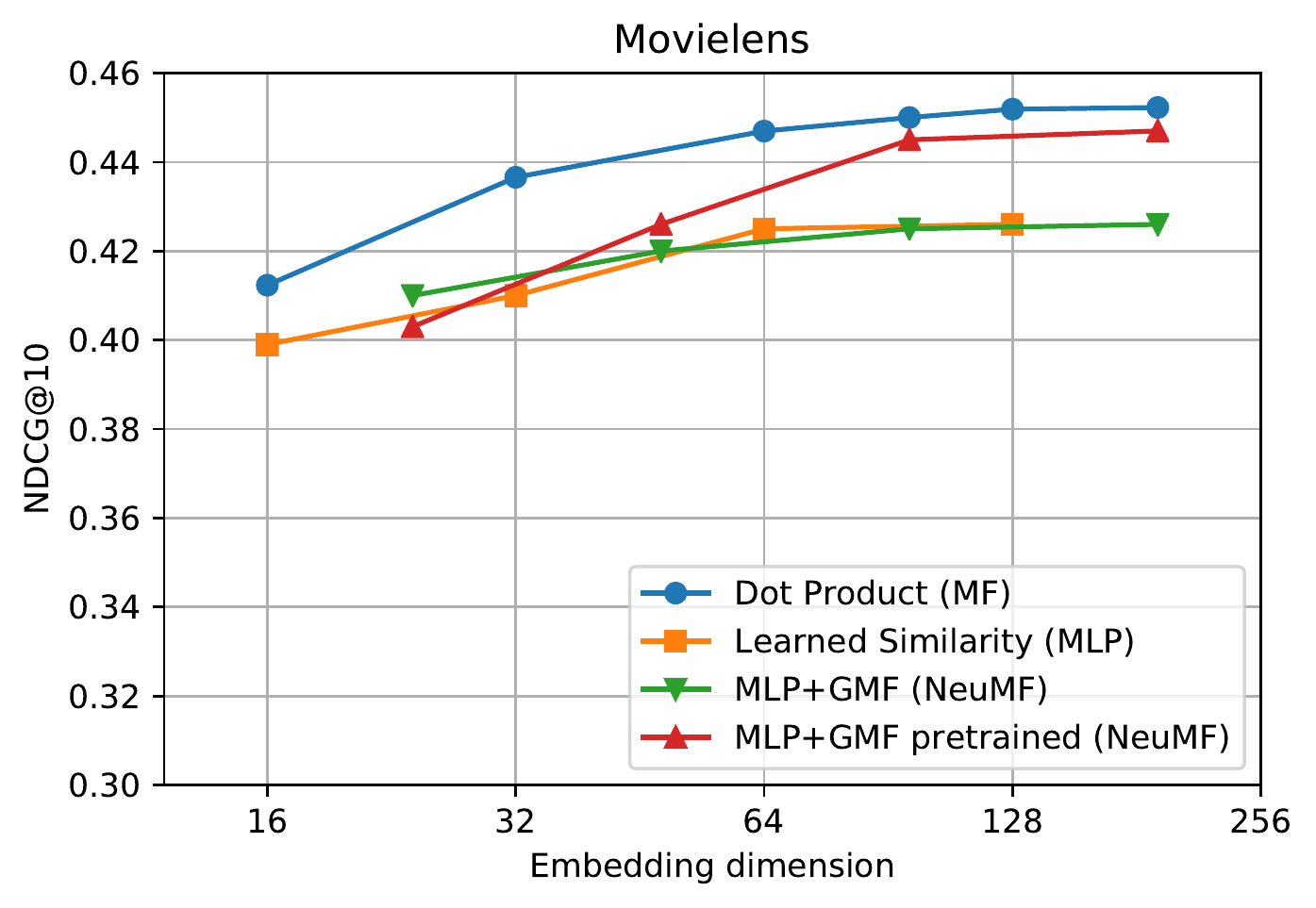}\\%
    \includegraphics[width=0.49\textwidth]{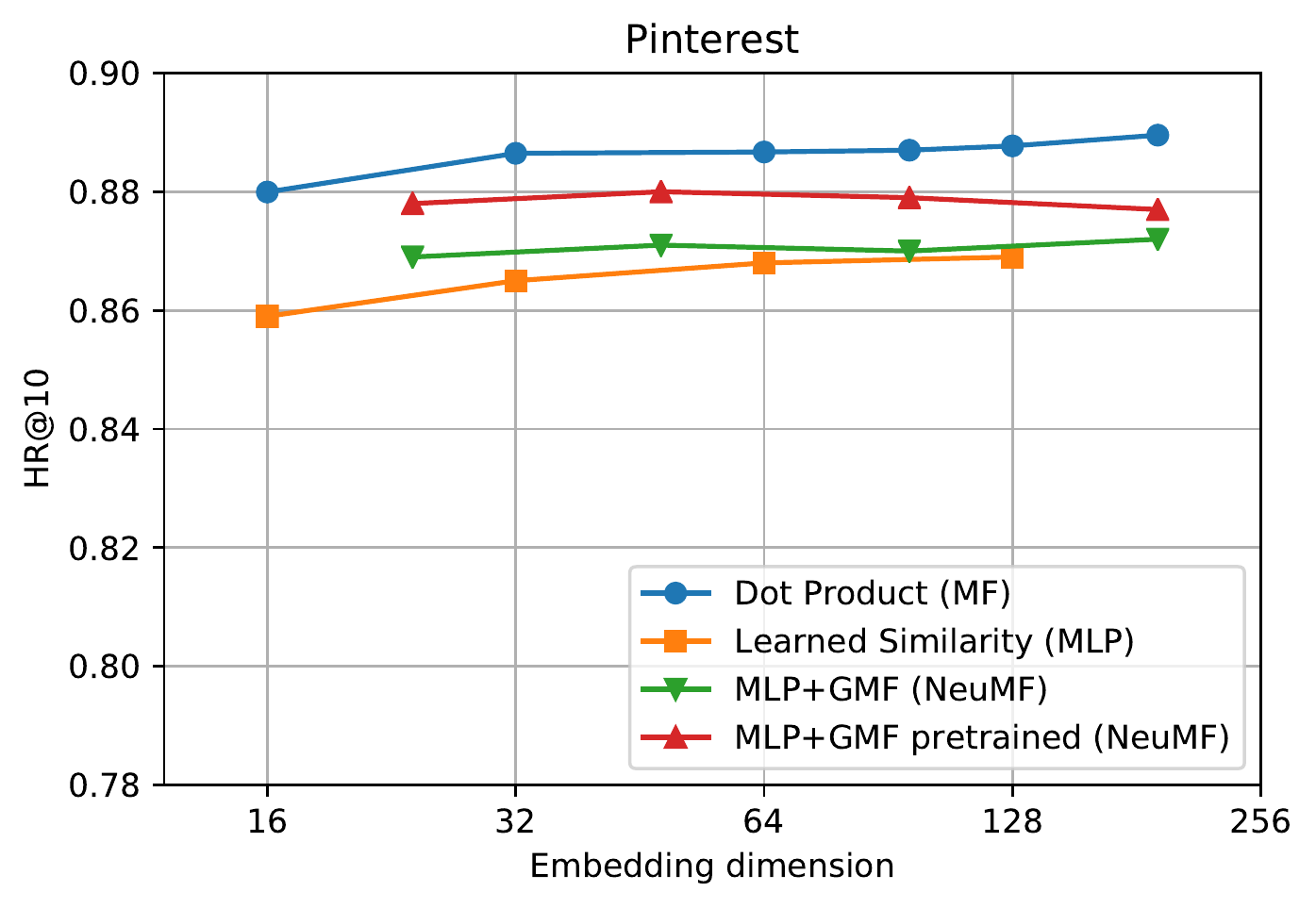}%
    \includegraphics[width=0.49\textwidth]{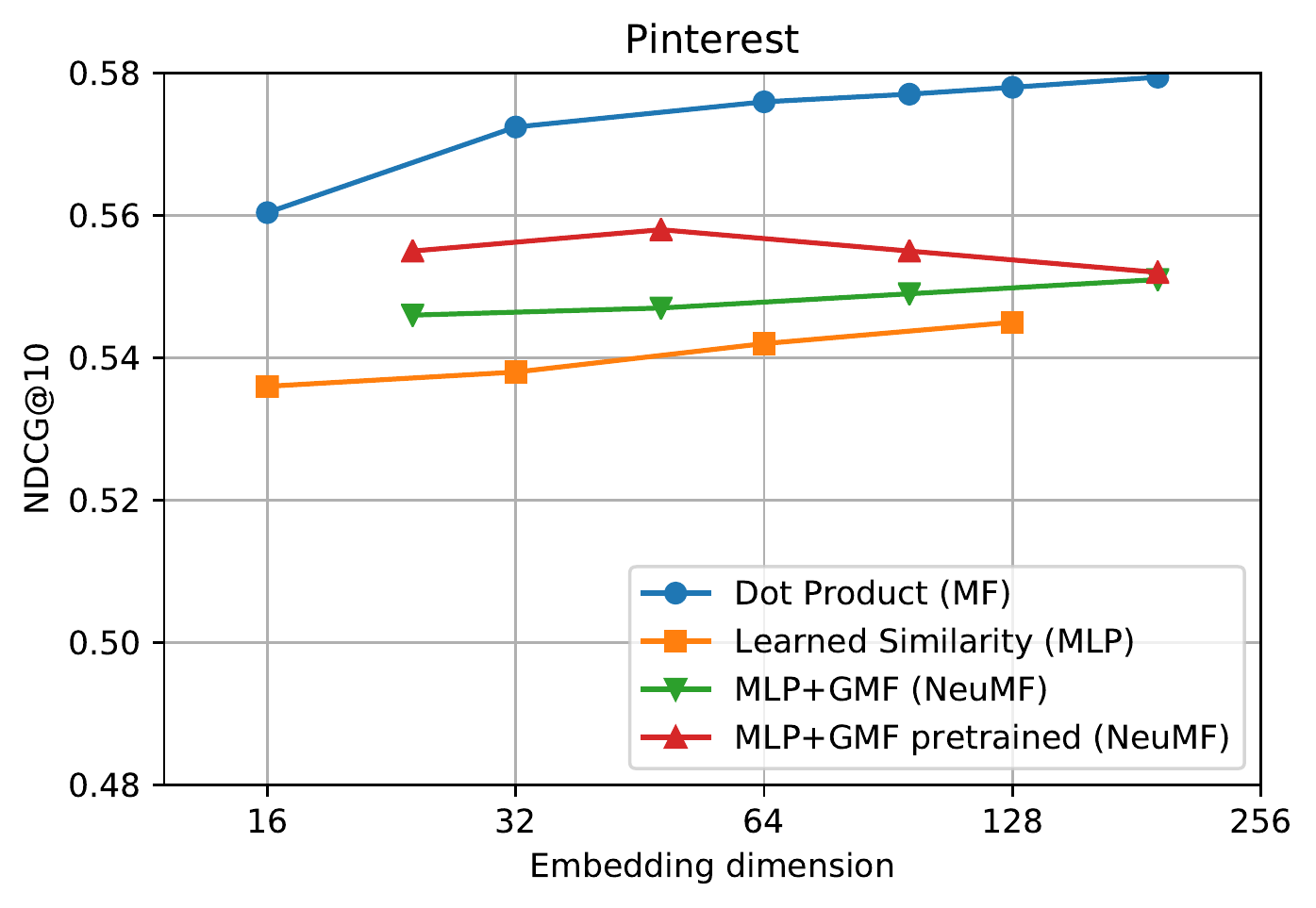}%
    \caption{Comparison of learned similarities (MLP, NeuMF) to a dot product: The results for MLP and NeuMF are from~\cite{he:www17}. The dot product substantially outperforms the learned similarity measures. Only the pretrained NeuMF is competitive, on one dataset, and for large embedding dimension.}
    \label{fig:quality}
\end{figure}

The results are reported in Fig.~\ref{fig:quality}.
Contrary to the findings of the NCF paper, the simple matrix factorization model exhibits the best quality over all evaluation metrics, and all embedding dimensions but one.

\subsubsection{Matrix Factorization vs MLP}
Our main interest is to investigate if the MLP-learned similarity is superior to a simple dot product.
As can be seen in Fig.~\ref{fig:quality}, the dot product substantially outperforms MLP on all datasets, evaluation metrics and embedding dimensions.
With a properly set up matrix factorization model, the experiments do not show any evidence that a MLP is superior.
In addition to a lower prediction quality, MLP-learned similarity suffers from other disadvantages compared to dot-product: the model has more model parameters (see Section~\ref{sec:learning_dot}), and is more expensive to serve (see Section~\ref{sec:serving}).

\subsubsection{Matrix Factorization vs NeuMF}

The NCF paper~\cite{he:www17} also proposes a combined model where the similarity function is a sum of dot-product and MLP, as in Eq.~\eqref{eq:ncf} -- this is called NeuMF\footnote{Following~\cite{he:www17}, the NeuMF uses 2/3rds of the embeddings for the MLP and 1/3rd for the MF. See the discussion about ``predictive factors" in Section~\ref{sec:predictive_factor} for details.}.
The green curve in Fig.~\ref{fig:quality} shows the performance of this combined model.
One can observe only a minor improvement over MLP and overall a much worse quality than MF.
The experiments do not support the claim in~\cite{he:www17} that a dot product model can be enhanced by feeding some part of its embeddings through an MLP.

A second variant of NeuMF was proposed in~\cite{he:www17}, that first trains MLP and MF models separately, then fine tunes the combined model. This can be viewed as a form of ensembling.
The red curve shows this variant, which performs better than training the combined model directly (in green), but performs worse than the MF baseline overall, except on one datapoint (HR on Movielens with embedding dimension $d = 192$).
Once again, the results do not support the claim that a learned similarity using a MLP is superior to a dot product. The experiment only indicates that ensembling two models can be helpful, a fact that has been observed for a variety of applications, and it is possible that ensembling with different models may yield a similar improvement. The fact remains that using a simple dot product outperforms this ensemble.

\subsubsection{On the performance of GMF}

Other variants of matrix factorization were considered in~\cite{he:www17}. In particular, the GMF model uses a weighted dot product $\phi^{\text{GMF}}$ as described in Eq.~\eqref{eq:gmf}. Except for the weights in the dot product, this model is very similar to the MF baseline we trained, in particular, both models use the same loss and negative sampling method.
Nevertheless, the GMF results reported in \cite{he:www17} are much worse than our MF results.
This discrepancy may seem surprising at first glance. We can see two reasons for this difference.
First, properly setting up and tuning baseline methods can be difficult in general, as argued in~\cite{rendle:arxiv19}, and the reported results may be improved by a more careful setup.

Second, $\phi^{\text{GMF}}$ introduces new model parameters -- the vector $\bw$ in Eq.~\eqref{eq:gmf}.
While this appears to be an innocuous generalization of the dot product similarity, it can have negative effects.
For example, L2 regularization of the embeddings ($\embu$ and $\embi$) is meaningless unless $\bw$ is regularized as well.
More precisely, suppose the loss function is of the form
\[
L(\Embu, \Embi, \bw, \lambda) = \ell(\{\phi^{\text{GMF}}_\bw(\embu, \embi) : \embu \in \text{Rows}(\Embu), \ \embi \in \text{Rows}(\Embi)\}) + \lambda (\|\Embu\|_F^2 + \|\Embi\|_F^2)
\]
where $\Embu, \Embi$ are embedding matrices, the first term of the loss $\ell$ depends on the pairwise similarities (i.e. the model output), and the second term is a regularization term, where $\Embu, \Embi$ are regularized but $\bw$ is not. Observe that if we scale the model parameters as $\Embu/a, \Embi/a, a^2\bw$ for some positive scalar $a$, then the model output is unchanged (given the expression of $\phi^{\text{GMF}}$), and we have
\begin{align}
    L(P,Q,\bw,\lambda)
    = L\left(\frac{1}{a}P,\frac{1}{a}Q,a^2\bw,a^2\lambda\right).
\end{align}
It follows that minimizing $L$ with a given $\lambda$ is equivalent to minimizing $L$ with any other $\tilde\lambda$ up to the change of variable $(P/a,Q/a,a^2\bw)$ with $a =\sqrt{\frac{\tilde{\lambda}}{\lambda}}$, a change of variable which leaves the model output unchanged. The solution is therefore unaffected by regularization. A second consequence is that unless $\lambda = 0$, minimizing the loss $L$ will likely result in embedding matrices $P, Q$ of vanishing norm and a vector of weights $\bw$ of diverging norm, leading to numerical instability.

The GMF results in~\cite{he:www17} support that the model is indeed not properly regularized because its results do not improve with a higher embedding dimension -- unlike in our experiments.

Finally, we observe that GMF does not improve model expressivity compared to a simple dot product, since the weights $\bw$ can simply be absorbed into the embedding matrices $P$ and $Q$.
This is another indicator that adding parameters to a simple model is not always a good idea and has to be done carefully.

\subsection{Further comparison}
\label{sec:non_cherry_picked}

As reported in the meta study of \cite{dacrema:arxiv2019}, the results for NeuMF and MLP in~\cite{he:www17} were cherry-picked in the following sense: the metrics are reported for the best iteration \emph{selected on the test set}.
The NeuMF and MLP numbers we report in Fig.~\ref{fig:quality} are from the original paper and likely over-estimate the actual test performance of those methods.
On the other hand, our MF results in Fig.~\ref{fig:quality} are not cherry picked, because we select all hyperparameters including the stopping iteration on a validation set -- see Appendix~\ref{sec:ncf_exp_details} for details.
The fact that our baseline MF outperforms the MLP-learned similarity despite the cherry-picking in the latter strengthens our conclusions.

In this section, we give an additional comparison using non cherry-picked results produced by~\cite{dacrema:arxiv2019}.
Table~\ref{tbl:dacrema} includes their results together with our matrix factorization (same as in Fig.~\ref{fig:quality}), with embedding dimension $d=192$.
The results confirm that the simple matrix factorization model substantially outperforms NeuMF on all metrics and datasets.
Our results provide further evidence to the conclusion of \cite{dacrema:arxiv2019} that simple, well-known baselines outperform NCF.
Note that matrix factorization was also one of the baselines in~\cite{dacrema:arxiv2019}  (the iALS method in Table~\ref{tbl:dacrema}), but our experiment shows a much larger margin than was obtained in~\cite{dacrema:arxiv2019}.

\begin{table}
\caption{Comparison from \cite{dacrema:arxiv2019} of MLP+GMF (NeuMF) with various baselines and our results. The best results are highlighted in bold, the second best result is underlined. \label{tbl:dacrema}}
\begin{tabular}{|l|rr|rr|r|}
    \hline
     Method                                     & \multicolumn{2}{c}{Movielens}     &  \multicolumn{2}{c}{Pinterest} & Result \\
                                               & HR@10     & NDCG@10   & HR@10     & NDCG@10 & from \\
    \hline
    Popularity                                  & 0.4535    & 0.2543    & 0.2740    & 0.1409 & \cite{dacrema:arxiv2019}\\
    SLIM~\cite{ning2011slim,levy2013efficient}  & \underline{0.7162}    & \underline{0.4468}    & 0.8679    & \underline{0.5601} & \cite{dacrema:arxiv2019}\\
    iALS~\cite{hu:icdm08}                       & 0.7111    & 0.4383    & 0.8762    & 0.5590 & \cite{dacrema:arxiv2019}\\
    MLP+GMF~\cite{he:www17}                     & 0.7093    & 0.4349    & \underline{0.8777}    & 0.5576 & \cite{dacrema:arxiv2019}\\
    Matrix Factorization
                                                & \textbf{0.7294} & \textbf{0.4523}    & \textbf{0.8895}    & \textbf{0.5794} & Fig.~\ref{fig:quality}\\
    \hline
\end{tabular}
\end{table}

\subsection{Discussion}

Following the arguments in~\cite{rendle:arxiv19}, it is possible that the studies in \cite{he:www17} and \cite{dacrema:arxiv2019} did not properly set up MLP and NeuMF, and that these results could be further improved.
It is also possible that the performance of these models is different on other datasets.
Nevertheless, at this point, the revised experiments from \cite{he:www17} provide no evidence supporting the claim that a MLP-learned similarity is superior to a dot product.
This negative result also holds for NeuMF where a GMF is added to the MLP.
And it also holds for the pretrained version of NeuMF.
Our study treats MLP and NeuMF favorably: (1)~we report the results for MLP and NeuMF that were obtained by the original authors, avoiding any bias in improperly running their methods.
(2)~These cited numbers for MLP and NeuMF are likely too optimistic as they were obtained through cherry picking as identified by \cite{dacrema:arxiv2019}.

\section{Learning a Dot Product with MLP is Hard}

An MLP is a universal function approximator: any continuous function on a compact set can be approximated with a large enough MLP~\cite{cybenko1989approximation,hornik1989multilayer,barron1993universal}.
It is tempting to argue that this makes the MLP a more powerful embedding combiner and it should thus perform at least as well or better than a dot product.
However, such an argument neglects the difficulty of learning the target function using MLPs: the larger class of functions also implies more parameters needed for representing the function.
Hence it would require more data to learn the function and may encounter difficulty in actually learning the desired target function.
Indeed, specialized structures, e.g. convolutional, recurrent, and attention structures, are common in neural networks.
There is probably no hope to replace them using an MLP though they should all be representable.
However, is this also true for the simple ``structure'' of the dot product?
Similar problems turn out to be actively studied subject in machine learning theory~\cite{andoni2014learning,li:nips17,du19c,allen-zhu19}.
To our knowledge, the best theoretical bound for learning the dot product, a degree two polynomial, requires $O(d^4/\epsilon^2)$ steps for an error bound of $\epsilon$~\cite{andoni2014learning}.
While the theory gives only a sufficient condition, it does hint that the difficulty scales polynomially with dimension~$d$ and $1/\epsilon$. This motivates us to investigate the question empirically.

\subsection{Experimental setup}
\label{sec:learning_dot}

We set up a synthetic learning task\footnote{The code is available at \url{https://github.com/google-research/google-research/tree/master/dot_vs_learned_similarity}.} where given two embeddings $\embu, \embi \in \mathbb{R}^d$ and a label $y(\embu, \embi)$, we want to learn a function  $\hat{y}: \mathbb{R}^{2\,d} \to \mathbb{R}$ that approximates $y$ with $\hat{y}(\embu,\embi)$.
We draw the embeddings $\embu, \embi$ from $\N(0,\sigma_{\text{emb}}^2 I)$ and set the true label as $y(\embu, \embi) = \langle \embu, \embi \rangle + \epsilon$ where $\epsilon \sim \N(0,\sigma_{\text{label}}^2)$ models the label noise.
From this process we create three datasets each consisting of tuples $(\embu, \embi, y)$.
One of the datasets is used for training and the remaining two for testing.
For the training and first test dataset, we first sample $M$ different user embeddings and $N$ different item embeddings, i.e., there are two fixed embedding matrices $\Embu \in \mathbb{R}^{M \times d}$ and $\Embi \in \mathbb{R}^{N \times d}$.
Then we uniformly sample (without replacement) $100 \, M$ user-item combinations and put 90\% into the training set and 10\% into the test set.
We create a second test set that consists of \emph{fresh} embeddings that did not appear in the training or test set, i.e., we sample the embeddings for every case from $\mathcal{N}(0,\sigma^2_\text{emb} I)$ instead of picking them from $\Embu$ and $\Embi$.
The motivation for this setup is to investigate if the learned similarity function generalizes to embeddings that were not seen during training.

We train the MLP on the training dataset and evaluate it on both test datasets. For the architecture of the MLP, we follow the suggestions in the NCF paper: we use an input layer of size $2d$ consisting of the concatenation of the two embeddings, and 3 hidden layers with sizes $[4\,h,2\,h,h]$ where $h$ is a parameter, and use the ReLU as the activation function.
The NCF paper suggests to use $h=d/2$, we also experiment with $h=d$ and $h=2d$.
For $h=d$, the number of model parameters are about $18\,d^2$, so for example for d=8: 1,152 or d=64: 73,728 or for d=256: 1,179,648.
For optimization, we also follow the NCF paper and choose the Adam optimizer.

As evaluation metric, we compute the RMSE between the predicted similarity of the MLP and the true similarity $y$.
We also measure the RMSE of a trivial model that predicts always 0 (=average rating in our dataset).
In our setup, this RMSE is equal in expectation to $\sqrt{\text{Var}(y)}=\sqrt{\sigma_{\text{label}}^2 + d\,\sigma_{\text{emb}}^4}$.
Secondly, we measure the RMSE of the dot product model, i.e., $\hat{y}(\embu, \embi)=\langle \embu, \embi \rangle$.
This RMSE is equal in expectation to $\sigma_{label}$.
We report the approximation error, i.e., the difference between the RMSE of the dot product model and the MLP.
Each experiment is repeated 5 times and we report the mean.

We want to choose the experimental parameters $\sigma_{label}$ and $\sigma_{emb}$ such that the approximation error gives some indication what values are acceptable.
To do this we choose values that are related to well-studied rating prediction tasks.
In the Netflix prize, the best models have RMSEs of about 0.85 \cite{koren:grandprize} -- for Movielens 10M, the best models have about 0.75 \cite{rendle:arxiv19}.
For these datasets, it is likely that the label noise is close to these values, thus we choose the label noise $\sigma_{label} = 0.85$.
For the Netflix prize, the trivial model that predicts always the average rating has an RMSE of 1.13.
Thus we set $\sigma_{emb}^2 = \sqrt{\frac{1.13^2 - 0.85^2}{d}}$.
With this setup, the trivial model in our experiment has the same RMSE as the trivial model on Netflix.
By aligning both the trivial model and the noise to the Netflix prize, absolute differences in our experiment give some indication of the scale of acceptable errors.
In both Netflix and ML 10M, a difference in RMSE of 0.01 is considered very large.
For example, for the Netflix prize it took the community about a year\footnote{\url{https://www.netflixprize.com/leaderboard_quiz.html}} to lower the RMSE from 0.8712 to 0.8616.
Similarly, for Movielens 10M, it took about 4 years  to lower the RMSE from 0.7815 to 0.7634.
Much smaller differences have been published. For example many published increments on Movielens 10M are about 0.001~\cite{rendle:arxiv19}.
We will use these thresholds of 0.01 and 0.001 as an indication whether the approximation errors are acceptable in our experiments.
While this is not a perfect comparison, we hope that it can serve as a reasonable indicator.

\subsection{Results}

\begin{figure}
    \centering
    \includegraphics[width=0.99\textwidth]{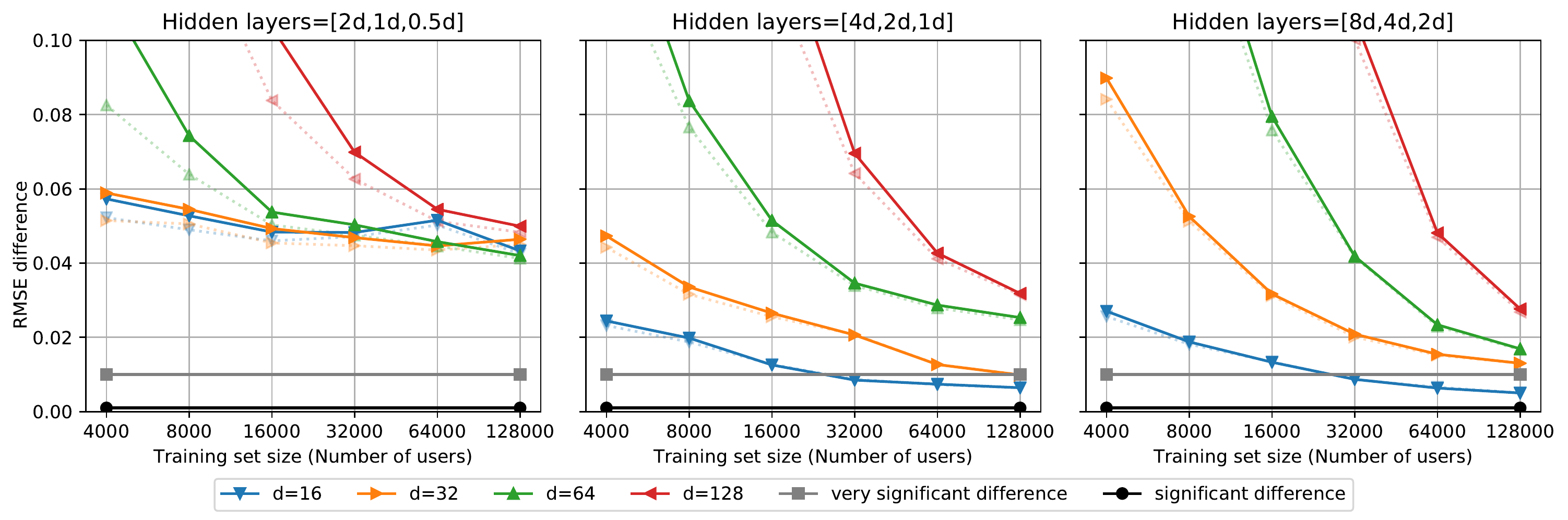}
    \caption{How well a MLP can learn a dot product over embeddings of dimension~$d$.
    The ground truth is generated from a dot product of Gaussian embeddings plus Gaussian label noise.
    The graphs show the difference between the RMSE of the dot product and the RMSE of the learned similarity measure; the solid line measures the difference on the fresh set, the dotted on the test set.
    Noise and scale have been chosen such that $0.01$ could indicate a very significant difference and $0.001$ a significant difference.}
    \label{fig:approx_dot}
\end{figure}

Figure~\ref{fig:approx_dot} shows the approximation error of the MLP for different choices of embedding dimensions and as a function of training data.
The figure suggests that with enough training data and wide enough hidden layers, an MLP can approximate a dot product.
This holds for embeddings that have been seen in the training data as well as for fresh embeddings.
However, consistent with the theory, the number of samples needed scales polynomially with the increasing dimensions and reduced error. Anecdotally, we observe the number of samples needed is about $O(d/\epsilon)^{\alpha}$ for $1\leq \alpha \leq 2$.
The experiments clearly indicate that it becomes increasingly difficult for an MLP to fit the dot product function with increasing dimensions.
In all cases, the approximation error is well above what is considered a large difference for problems with comparable scale. For example, for the moderate $d=128$, with $128000$ users, the error is still above $0.02$, much higher than the \emph{very significant} difference of $0.01$.

This experiment shows the difficulty of using an MLP to approximate the dot product, even when explicitly trained to do so.
Hence, if the dot product performs well on a given task, there could be a significant price to pay for an MLP to approximate it.
We hope this can explain, at least partially, why the dot product model outperforms the MLP model in the experiments of Section~\ref{sec:ncf_exp}.

\section{Applicability of Dot Product Models}
\label{sec:serving}

Most academic studies focus on training runtime when discussing applicability.
However, in industrial applications, the serving runtime is often more important, in particular when the recommendations cannot be precomputed offline but need to be computed at the time of the user's request. This is the case for most context-aware recommenders in which the recommendation depends on contextual features that are only available at query time.
For instance, consider a sequential recommender that recommends items to a user based on the previously selected L items.
Here the top scoring items cannot be precomputed for all possible combinations of L items.
Instead the recommender would need to retrieve the highest scoring items from the  whole item catalogue with a latency of a few milliseconds after the user's request.
Such real time retrieval is a common application in real world recommender systems \cite{covington:rs16}.

Computing a dot product similarity takes $\O(d)$ time while computing an MLP-learned similarity takes $\O(d^2)$ time.
If there are $n$ items to score, then the total costs are $\O(d n)$ (for dot) vs $\O(d^2 n)$ (for MLP).
For large scale applications, $n$ is typically in the range of millions and $d$ is in the hundreds, and while dot has a lower complexity, both are impractical for retrieval applications that require latencies of a few milliseconds.
However, for a dot product, the problem of finding the top scoring items can be approximated efficiently. Indeed, given the user embedding $\embu$, the problem is to find items $i$ that maximize $\langle \embu, \embi_i\rangle$. This is a well-studied problem, known as \emph{approximate nearest neighbor search} \cite{liu:nips04} or \emph{maximum inner product search} \cite{shrivastava:nips14}.
Efficient sublinear time algorithms exist that makes dot product retrieval feasible in typically a few milliseconds, even with millions of items $n$ \cite{covington:rs16}.
To the best of our knowlegde, no such sublinear techniques exist for nearest neighbor retrieval with MLPs.

To summarize, MLP similarity is not applicable for real time top-N recommenders, while the dot product allows fast retrieval using well established nearest neighbor search algorithms.

\section{Related Work}

\subsection{Dot products at the Output Layer of DNNs}

\label{sec:dnn}
At first glance it might appear that our work questions the use of neural networks in recommender systems.
This is not the case, and as we will discuss now, many of the most competitive neural networks use a dot product for the output but not an MLP.
Consider the general multiclass classification task where $(\bx,y)$ is a labeled training example with input $\bx$ and label $y \in \{1,\ldots,n\}$.
A common approach is to define a DNN $\textbf{f}$ that maps the input $\bx$ to a representation (or embedding) $\textbf{f}(\bx) \in \mathbb{R}^d$.
At the final stage, this representation is combined with the class labels to produce a vector of scores.
Commonly, this is done by multiplying the input representation $\textbf{f}(\bx) \in \mathbb R^d$ with a class matrix $\Embi \in \mathbb{R}^{n \times d}$ to obtain a scalar score for each of the $n$ classes.
This vector is then used in the loss function, for example as logits in a softmax cross entropy with the label $y$.
This falls exactly under the family of models discussed in this paper, where $\embu = \textbf{f}(\bx) \in \mathbb{R}^d$ and the classes are the items.
In fact, the model as described above is a dot product model because at the output $\Embi\,\textbf{f}(\bx) = \Embi\,\embu = [\langle \embu, \embi_i \rangle]_{i=1}^n$ which means each input-label or user-item combination is a dot product between an input (or user) embedding and label (or item) embedding.
This dot product combination of input and class representation is commonly used in sophisticated DNNs for image classification~\cite{alexnet-2012,he:cvpr16} and for natural language processing~\cite{bengio2003neural,mikolov2013distributed,devlin:arxiv18}.
This makes our findings that a dot product is a powerful embedding combiner well aligned with the broader DNN community where it is common to apply a dot product at the output for multiclass classification.

\subsection{MLPs at the Output Layer of DNNs}

NeuMF is very closely related to the previously proposed \emph{neural network matrix factorization}~\cite{dziugaite:arxiv15}.
Neural network matrix factorization also uses a combination of an MLP plus extra embeddings with an explicit dot product like structure as in GMF.
A follow up paper \cite{he:ijcai18} proposes to replace the MLP in NCF by an outerproduct and pass this matrix through a convolutional neural network.
Finding the dot product with this technique is trivial because the sum of the diagonal in the outerproduct is the dot product.
Unfortunately, while written by the same authors as the NCF paper, it evaluates on different data, so our results in Section~\ref{sec:ncf_exp} cannot be compared to their numbers and it remains unclear if their work improves over a well tuned baseline with a dot product.
Besides prediction quality, this proposal suffers from the same applicability issues as the MLP (see Section~\ref{sec:serving}).

\subsection{Specialized Structures inside a DNN}
\label{sec:related:dnn}

In DNN modeling it is very common to replace an MLP by a more specialized structure that has an inductive bias that represents the problem better.
For example, in image classification structures such as convolutional neural networks are very popular because they represent the spatial structure of the input data.
In recurrent neural networks, such parameter sharing is very important too.
Another example are attention models, e.g. in Neural Machine Translation~\cite{wu2016google} and in the Transformer model~\cite{vaswani2017attention}, that contain a matrix product inside the neural network for combining multiple inputs -- they can be regarded as the dot product model for combining ``internal'' embeddings too.
All these specialized structures are crucial for advancing the state of the art of deep learning, although in theory they can all be approximated by MLPs.

The inefficiency of MLPs to capture dot and tensor products has been studied by \cite{beutel:wsdm18} in the context of recommender systems.
Here the authors examine how to add context to recurrent neural networks.
Similar to our work and Section~\ref{sec:learning_dot}, \cite{beutel:wsdm18} points out that MLPs do not model multiplications and it investigates approximating dot products and tensor products with MLPs empirically.
Their study focuses on the model size required to learn a tensor product for embeddings of dimension $d=1$ and $d=2$, where the number of distinct embeddings is 100 per mode and the training error is measured.

\subsection{Experimental Issues in Recommender Systems}

In their meta study, \cite{dacrema:arxiv2019} point out issues with evaluation in recommender system research.
Their experiments also cover the NCF paper.
They show that well studied baselines can get comparable results to (a reproducible value of) NeuMF (see Section~\ref{sec:non_cherry_picked}).
The goal of our study and \cite{dacrema:arxiv2019} is different.
While \cite{dacrema:arxiv2019} covers a broad set of methods and publications, we are investigating the specific issue of learned similarity functions in more detail.
Our work provides apples to apples comparisons of dot product vs MLP, stronger results (outperforming the original NCF results), and a thorough investigation of the reasons and consequences.

\section{Conclusion}

Our findings indicate that a dot product might be a better default choice for combining embeddings than learned similarities using MLP or NeuMF.
Shifting the focus in the recommender system research community from learned similarities to dot products might have several positive effects:
(1)~The research becomes more relevant for the industry because models are applicable (see Section~\ref{sec:serving}).
(2)~Dot product similarity simplifies modeling and learning (no pretraining, no need for large datasets) which facilitates both experimentation and understanding.
(3)~Better alignment with other research areas such as natural language processing or image models where the dot product is commonly used.

Finally, our experiments give further evidence that running machine learning methods properly is difficult~\cite{rendle:arxiv19} and one-off studies are prone to drawing wrong conclusions.
Introducing shared benchmarks might help to better identify improvements.

\appendix

\section{Experiments from NCF Paper}
\label{sec:ncf_exp_details}

This section provides details about our setup for establishing a dot product baseline for the experiments of the NCF paper (Section~\ref{sec:ncf_exp}).
The code and datasets of NCF were provided by its authors\footnote{\url{https://github.com/hexiangnan/neural_collaborative_filtering}}.
We provide code for our implementation of matrix factorization and the script to generate the tuning split\footnote{\url{https://github.com/google-research/google-research/tree/master/dot_vs_learned_similarity}}.

\subsection{Model and Optimization}
We implemented a matrix factorization with bias (see Eq.~\ref{eq:mfbias}).
The parameters of this model are the embeddings $\Embu \in \mathbb{R}^{M \times d}$ for $M$ users and $\Embi \in \mathbb{R}^{N \times d}$ for $N$ items.
Following the NCF paper, for training we cast the implicit data, which contains only positive observations, into a binary two class classification problem and sample $m$ negative items for each tuple in the implicit data.
In each epoch a new set of negatives is drawn -- the sampling distribution is uniform.
We minimize the binary logistic loss with L2 regularization.
For each training example $(u,i,y)$ where $y \in \{0,1\}$ is the binary label, the regularized loss is
\begin{align}
    l(u,i,y) = -y\,\ln \sigma(\phi(\embu_u,\embi_i)) - (1-y)\,\ln (1-\sigma(\phi(\embu_u,\embi_i))) + \lambda\,\|\embu_u\| + \lambda\,\|\embi_i\|
\end{align}
with the regularization constant $\lambda \in \mathbb{R}^+$.
The loss is optimized with stochastic gradient descent with learning rate $\eta$, with the update rules:
\begin{align}
    \sembu_{u,1} &\leftarrow \sembu_{u,1} - \eta [(\sigma(\phi(\embu_u,\embi_i)) - y)  + \lambda \sembu_{u,1}] \\
    \sembi_{i,1} &\leftarrow \sembi_{i,1} - \eta [(\sigma(\phi(\embu_u,\embi_i)) - y)  + \lambda \sembi_{i,1}] \\
    \embu_{u, [2,\ldots,d]} &\leftarrow \embu_{u, [2,\ldots,d]} - \eta [(\sigma(\phi(\embu_u,\embi_i)) - y) \, \embi_{i, [2,\ldots,d]} + \lambda \embu_{u, [2,\ldots,d]}] \\
    \embi_{i, [2,\ldots,d]} &\leftarrow \embi_{i, [2,\ldots,d]} - \eta [(\sigma(\phi(\embu_u,\embi_i)) - y) \, \embu_{u,[2,\ldots,d]} + \lambda \embi_{i, [2,\ldots,d]}]
\end{align}
The embeddings are initialized from a normal distribution.
This configuration shares the same loss, regularization, negative sampling approach, and initialization procedure with MLP and NeuMF as proposed in~\cite{he:www17}.

The hyperparameters of the dot product model are: embedding dimension $d$, regularization $\lambda$, learning rate $\eta$, number of negative samples $m$, number of training epochs, standard deviation for initialization.
Analogously to the NCF paper, we report results for $d \in \{16,32,64,96,128,192\}$.

\subsection{Hyperparameter Tuning}

We create a tuning dataset that follows the same splitting protocol as the final training/test split.
In particular, we remove the last feedback from each user from the training set and place it in a test set for tuning and keep the remaining training cases in a training set for tuning.
We then train models on the training set for tuning and evaluate the model on the test set for tuning.
We choose all hyperparameters including the number of training epochs on this tuning set.
Note that both the training set for tuning and test set for tuning contain no information about the final test set.

From our past experience with matrix factorization models, if the other hyperparameters are chosen properly, then the larger the embedding dimension the better the quality -- our experiments Figure~\ref{fig:quality} confirm this.
For the other hyperparameters: learning rate and number of training epochs influence the convergence curves.
Usually, the lower the learning rate, the better the quality but also the more epochs are needed.
We set a computational budget of up to 256 epochs and search for the learning rate within this setting.
In the first hyperparameter pass, we search a coarse grid of learning rates $\eta \in \{0.001, 0.003, 0.01\}$ and number of negatives $m = \{4,8,16\}$ while fixing the regularization to $\lambda = 0$.
Then we did a search for regularization in $\{0.001, 0.003, 0.01\}$ around the promising candidates.
To speed up the search, these first coarse passes were done with 128 epochs and a fixed dimension of $d=64$ (Movielens) and $d=128$ (Pinterest).
We did further refinements around the most promising values of learning rate, number of negatives and regularization using $d=128$ and 256 epochs.
Throughout the experiments we initialize embeddings from a Gaussian distribution with standard deviation of $0.1$; we tested some variation of the standard deviation but did not see much effect.

The final hyperparameters for Movielens are: learning rate $\eta=0.002$, number of negatives $m=8$, regularization $\lambda=0.005$, number of epochs 256.
For Pinterest: learning rate $\eta=0.007$, number of negative samples $m=10$, regularization $\lambda=0.01$, number of epochs 256.

After hyperparameter selection, we trained on the full dataset with these hyperparameters and evaluated according to the protocol in \cite{he:www17}.
We repeated the final training and evaluation 8 times and report the mean of the metrics.

\subsection{MLP and NeuMF Results}

We report the results for MLP and NeuMF from the original NCF paper~\cite{he:www17}.
As we share the same evaluation protocol and splits, the numbers are comparable.
We report the results for NeuMF from Table~2 in \cite{he:www17} and the results for MLP from Tables~3,4 in \cite{he:www17} using the `MLP-3' setting.

\label{sec:predictive_factor}
It should be noted that in \cite{he:www17}, the tables and plots use ``predictive factor" instead of embedding dimension. The predictive factor is defined as the size of the last hidden layer of the MLP, and as described in~\cite{he:www17}, for the 3-layer MLP a predictive factor of $k$ operates on two input embeddings, each of dimension $d=2\,k$.
For the NeuMF model, a predictive factor of $k$ operates on embeddings of dimension $d=3\,k$ because it consists of an independent MLP with predictive factor of $k$ (embedding size $d=2\,k$) and a GMF with embedding size $d=k$.
This definition of \emph{predictive factor} is arbitrary, in fact it can be made arbitrarily small by adding layers to the MLP without changing anything else in the model.
We think it is more meaningful to compare models with a fixed embedding dimension.
In particular, we want to investigate the prediction quality of an MLP or a dot product over two embeddings of the same size $d$.
We recast all results from the NCF paper in terms of embedding dimension, by multiplying the predictive factor by 3 for NeuMF results and by~2 for MLP results.
This allows us to do an apples to apples comparison of different similarity functions over an embedding space of dimension $d$.

\end{document}